# From Black Holes to Cosmology : The Universe in the Computer

**J.-P. Luminet**
Laboratoire Univers et Théories (LUTH)
Observatoire de Paris, CNRS, Université Paris Diderot
5 Place Jules Janssen, 92190 Meudon, France

E-mail: jean-pierre.luminet @obspm.fr

**Abstract.** I discuss some aspects of the use of computers in Relativity, Astrophysics and Cosmology. For each section I provide two examples representative of the field, including gravitational collapse, black hole imagery, supernovae explosions, star-black hole tidal interactions, N-body cosmological simulations and detection of cosmic topology.

**1. Introduction**
In this plenary lecture I shall discuss various aspects of the use of computers in relativity, astrophysics and cosmology, by presenting a few representative examples of what is been done in the field. These examples are by no means exhaustive, but reflect a personal choice on subjects that I know better than others, such as black holes and cosmology, and I apologize for not discussing all the excellent work in numerical astrophysics that could not find place in this short presentation.

Of course the use of computers is present in all the activities of astronomers and astrophysicists, but it is important to distinguish between numerical modelisation (involving computer simulations) from data treatment and analysis – although both may consume a lot of computing time and memory. As a theoretician, I shall concentrate mainly on numerical modelisation.

Before discussing the physics, let us begin with a brief survey of the computing facilities available to the astronomical community, from the local scale of a laboratory to the global scale of international organizations.

• At the local scale, e.g. a typical laboratory comprising about fifty researchers in theoretical astrophysics, such as mine (Laboratory Universe and Theories), most of calculations are performed on sequential machines – personal laptops and computers - and small clusters (10 to 100 nodes) owned by the laboratory. For heavier calculations, use is done of mesocenters (around 200 nodes) located in Paris Observatory, grids and supercomputers (more than 10,000 nodes) owned by national institutions. Numerical tools are developed, including codes for hydrodynamics, shock capture methods, N-body simulations, ray-tracing, etc., as well as numerical libraries, which provide a set of tools in the form of free public softwares for building various codes. For instance the publicly available library Kadath [http://luth.obspm.fr/~luthier/grandclement/kadath.html] implements spectral methods for solving partial differential equations in general relativity, astrophysics and theoretical physics, written in C++ or Fortran 77. Fully parallel and running on several hundreds of processors, Kadath has successfully recovered known solutions in various areas of physics and is the main tool for the study of quasi-periodic solutions in gauge field theory [1].

• At the scale of a country such as France, computing facilities are provided by governmental organizations. For instance, the GENCI (Grand Equipement National de Calcul Intensif [http://www.genci.fr/en]) coordinates the principal French equipments in high performance computing. This legal entity takes the form of a «société civile» under French law, owned by the French State represented by the Ministry for Higher Education and Research, by CEA (Commission for alternative Energies and Atomic energy), CNRS (National Centre for Scientific Research) and Universities. Created in 2007, GENCI implements and ensures the coordination of the major equipments of the national High Performance Computing (HPC) centres by providing funding and by assuming ownership, promotes the organization of an European HPC area and participates to its achievements, and sets up R&D collaborations in order to optimize HPC.

• At a more global scale such as European Union, the PRACE infrastructure (Partnership foR Advanced Computing in Europe, [http://www.prace-ri.eu]) creates a pan-European supercomputing infrastructure for large scale scientific and engineering applications. The computer systems and their operations accessible through PRACE are provided by 4 country members (BSC representing Spain, CINECA representing Italy, GCS representing Germany and GENCI representing France) who committed a total funding of €400 million for the initial PRACE systems and operations. In pace with the needs of the scientific communities and technical developments, systems deployed by PRACE are continuously updated and upgraded to be at the apex of HPC technology.

The Curie supercomputer, owned by GENCI and operated by CEA, is the first French Tier0 system open to scientists through the French participation into the PRACE infrastructure. It offers a peak performance of 1,7 Petaflops but at the present date (september 2013) it ranks only #15 in the list of the most powerful computers.

For the sake of comparison, table 1 below lists some of the TOP 20 supercomputer sites for June 2013 (keeping in mind that the list evolves rapidly).

**Table 1.**

| Rank | Site | Name | Cores | Rpeak *(Petaflops)* |
|---|---|---|---|---|
| 1 | National University of Defense Technology (China) | Tianhe 2 | 3,120,000 | 55 |
| 2 | DOE/SC/Oak Ridge National Lab (United States) | Titan | 560,640 | 27 |
| 3 | DOE/NNSA/LLNL (UnitedStaes) | Sequoia | 1,572,864 | 20 |
| 4 | RIKEN/AICS (Japan) | K computer | 705,024 | 11 |
| 5 | DOE/SC/Argonne National Lab (United States) | Mira | 786,432 | 10 |
| ... | ... | ... | ... | ... |
| 7 | Forschungszentrum Juelich (Germany) | JUQUEEN | 458,752 | 5,8 |
| ... | ... | ... | ... | ... |
| 12 | CINECA (Italy) | Fermi | 163,840 | 2,1 |
| ... | ... | ... | ... | ... |
| 15 | CEA/TGCC-GENCI (France) | Curie | 77,184 | 1,7 |

We conclude this introduction by recalling the common definition of what is called a "numerical grand challenge" : a fundamental problem in science or engineering, with broad applications, whose solution would be enabled by the application of HPC resources that could become available in the near future. Each grand challenge requires simulations on supercomputers with more than 10,000 nodes. The specificites and requirements of such programs are massive parallelization, very large fluxes of I/O data, and full teams of researchers, engineers and technicians to correctly operate them. In the last section of this article we briefly describe the numerical grand challenge in cosmology.

## 2. Numerical Relativity

Numerical relativity aims to obtain solutions to Einstein's equations with computers. As it provides a powerful tool to explore fundamental problems in physics and astrophysics, it has been a constant field of research since the 1990's, but spectacular progress has been made in the last decade due to
- larger computational facilities
- more advanced and accurate numerical techniques
- new formulation of Einstein's and magnetohydrodynamics (MHD) equations well-suited for numerical evolution.

Indeed a very important step is the choice of formulation and gauge for Einstein's equations : as these are covariant, one must specify a "good" gauge in order to be able to write them as a well-posed system of partial differential equations, which can also be numerically integrated without the appearance of instabilities. In this section we discuss two representative examples, related to gravitational collapse and black hole imagery.

### 2.1. From gravitational collapse to gamma-ray bursts

Gamma-ray bursts (GRBs) are flashes of gamma rays associated with extremely energetic explosions that have been observed in distant galaxies. They can last from ten milliseconds to several minutes, and their light curves vary considerably from one event to another. Two relatively distinct groups can nevertheless be identified: short bursts (SGRB), representing about one-third of those observed, which have a mean duration of 0.3 seconds, and long bursts, which last more than 2 seconds, with a mean of 30 seconds. It is tempting to explain this apparent dichotomy by different physical origins, especially as short bursts generally emit harder radiation (i.e higher frequency) than long bursts. Most investigators consider that GRBs are linked to the formation of stellar mass black holes, whose stronger gravitational fields can result in the release of more energy. Stellar black holes could be formed in at least two ways: either by the catastrophic collapse of a very massive star, or by the fusion of two compact stars.

In the model of coalescence, bursts would be formed from a pair of neutron stars orbiting around each other or from a pair consisting of a neutron star and a black hole. The theory of general relativity indicates that in such a situation the two compact stars would rapidly lose orbital energy in the form of gravitational waves. Over time, the decrease in the energy of the pair will inexorably shorten the distance between them. The ballet will end when the two bodies collide and fuse, thus giving birth to a black hole and an accretion disc, accompanied by a spurt of ultrahot matter orthogonal to the disc, which is responsible for a short burst.

This qualitative scenario is supported to a good extent by fully general-relativistic simulations. I refer the reader to the excellent review by Rezzolla [2]. The numerical investigation of the inspiral and merger of binary neutron stars in full general relativity has seen enormous progress made in recent years. Crucial improvements in the formulation of the equations and numerical methods, along with increased computational resources, have extended the scope of early simulations. These developments have made it possible to compute the full evolution, from large binary-separations up to black-hole formation, without and with magnetic fields and with idealised or realistic equations-of-state. Numerical simulations show that the formation of a torus around a rapidly rotating black hole is inevitable (figure 1). They also provide the first evidence that the merger of a binary of modestly magnetised neutron stars naturally forms many of the conditions needed to produce a jet of ultrastrong magnetic field, with properties that are consistent with SGRB observations. This missing link between the astrophysical phenomenology of GRBs and the theoretical expectations is a genuine example of the potential of numerical relativity.

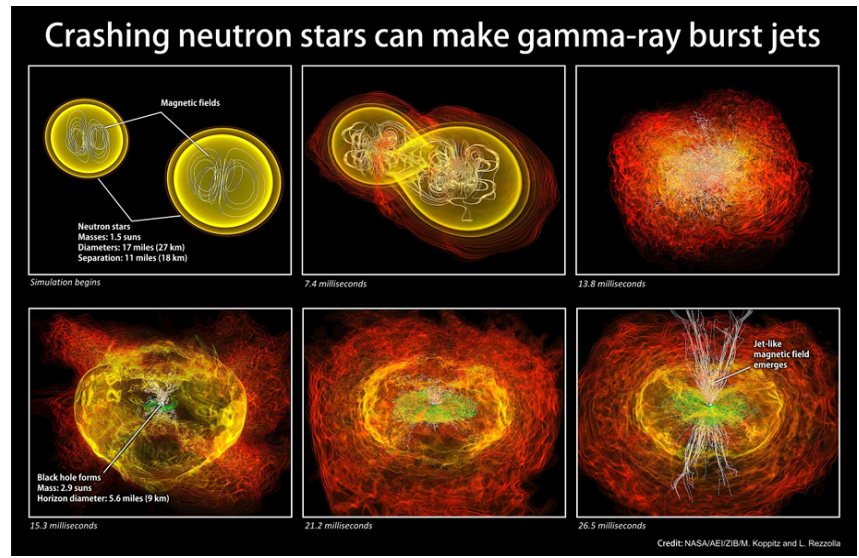

**Figure 1.** These images simulate the merger of two neutron stars using a new supercomputer model. Ribbons and lines represent magnetic fields. The orbiting neutron stars rapidly lose energy by emitting gravitational waves and merge in less than 8 milliseconds. The merger amplifies and scrambles the merged magnetic field. A black hole forms and the magnetic field becomes more organized, eventually producing structures capable of supporting the jets that power SGRB. Credit: NASA/AEI/ZIB/M. Koppitz and L. Rezzolla.

This remarkable advancement also provides information about the entire gravitational waveform, from the early inspiral up to the ringing of the black hole. Advanced interferometric detectors starting from 2014 (advanced Virgo /advanced LIGO) are expected to observe such sources at a rate of 40 – 400 events per year.

*2.2. Black hole imagery*
Since black holes cause extreme deformations of spacetime, they also create the strongest possible deflections of light rays passing in their vicinity and give rise to spectacular optical illusions. Today, it is possible to observe black holes only indirectly, through the effects that they have on their environment; for instance, their powerful gravitational field sucks in the neighboring gaz into accretion disks.

The first computer images of the appearance of a black hole surrounded by an accretion disk were obtained by myself in 1979 [3]. Calculations were done on an IBM 7040 machine of Paris Observatory; it was the time of punched cards, and no visualization device was available, so that I had to produce the final image by hand from numerical data (figure 2) ! The system is observed from a great distance at an angle of 10° above the plane of the disk. The light rays are received on a photographic plate (rather a bolometer in order to capture all wavelengths). A realistic image, e.g. taking account of the spacetime curvature, of the blue- and redshift effects, of the physical properties of the disk and so on, can be precisely calculated at any point of spacetime. Because of the curvature of spacetime in the neighborhood of the black hole, the image of the system is very different from the ellipses which would be observed if an ordinary celestial body (like the planet Saturn) replaced the black hole. The light emitted from the upper side of the disk forms a direct image and is considerably distorted, so that it is completely visible. There is no hidden part. The lower side of the disk is also visible as an indirect image, caused by highly curved light rays. In theory, there are multiples images of higher orders that give extremely distorted views of the top and of the bottom that are even more squashed, and so on to infinity.

With the improvement of computers, color and animated calculations were later performed by Marck [4] on a DEC-VAX 8600 computer, see fig. 3. At the time, black hole imagery was done mainly for pedagogical purpose, but the situation has changed very recently. Indeed black hole physics will develop considerably in the coming years, driven by observations at very high angular resolution (micro-arcseconds), as well in infrared (GRAVITY instrument installed at the VLT in 2014) as at submillimeter wavelength (Event Horizon Telescope, circa 2020). For the first time, we shall see the immediate environment of the event horizon of a black hole, especially the central black hole of our Galaxy, Sgr A*. Thus it is now essential to prepare the observations by numerical simulations which compute images and spectra of black holes and their close environment. Figure 4 depicts the result recently obtained by a new ray-tracing code, GYOTO (General relativitY Orbit Tracer of Paris Observatory) [5], publicly released at http://gyoto.obspm.fr. With respect to existing codes, GYOTO has the distinctive capability to compute geodesics in any metric, in particular in metrics that are known only numerically. A first application has been the modelling of the spectrum of Sgr A* by a magnetized ion torus [6], see figure 5.

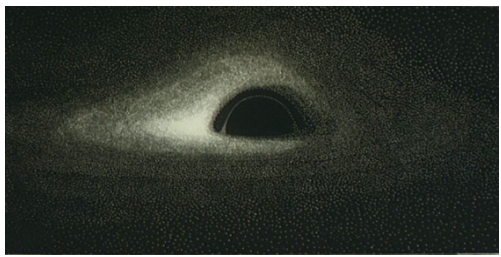

**Figure 2.** First simulation of a spherical black hole with thin accretion disk, 1979. Credit: J.-P. Luminet/CNRS

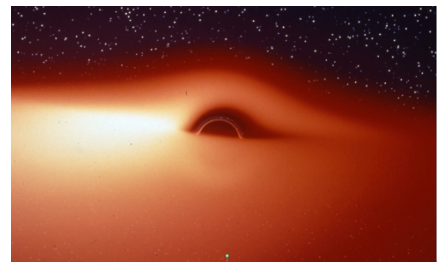

**Figure 3.** Image of a spherical black hole with thin accretion disk, 1996. Credit: J.-A. Marck/CNRS

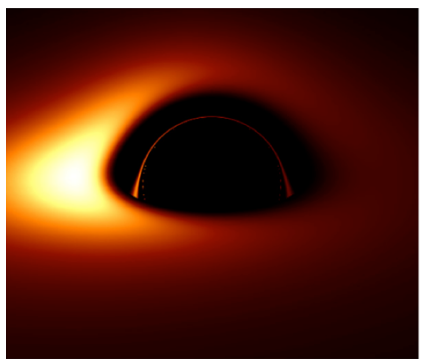

**Figure 4.** Same image calculated by a ray-tracing code, 2011. Credit: Vincent et al. /CNRS

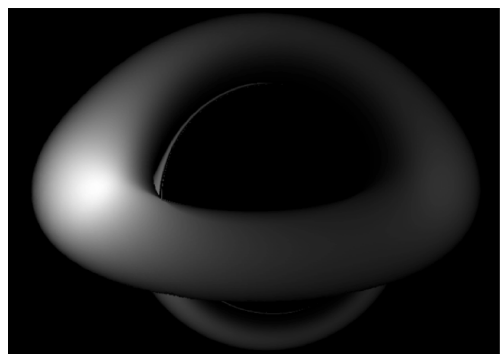

**Figure 5.** Image of an ion torus around a Kerr black hole, with radiative transfer, 2011. Credit: Vincent et al./CNRS

Another spectacular example of black hole imagery is the simulation of gravitational lensing. The spacetime curvature created by a massive object along the line of sight of a starry sky causes optical illusions for all objects in the background, because the curvature increases the number of distinct trajectories taken by light rays. Powerful modern telescopes can detect gravitational mirages created by intervening ordinary bodies such as stars (micro-lensing), galaxies or galaxy clusters which function as a gravitational lens, but not yet those created by black holes, due to lack of resolution. Computer simulations [7] recreate the effect of a large black hole acting as a gravitational lens to distort a starry landscape in the background and multiply the images of distant objects to create a mirage. Figures 6-7 show gravitational lensing and optical illusions produced by a black hole in the line of sight of the Milky Way center in the Southern Hemisphere (left) and in front of the Magellanic Clouds (right).

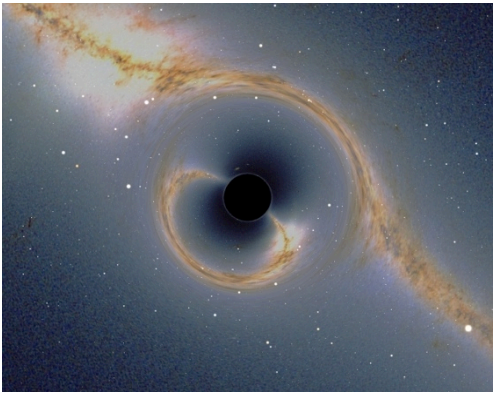
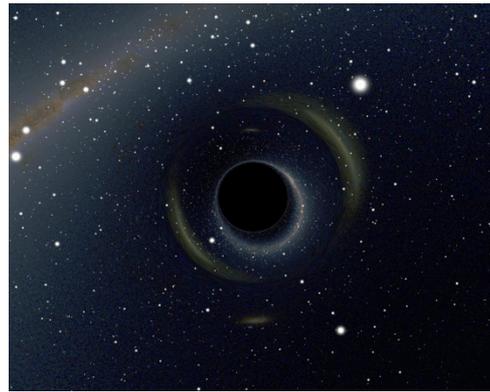

**Figure 6.** Simulation of gravitational lensing by a black hole in the Southern Hemisphere. The full Milky Way is seen as a mirror image close to the black hole. Credit A. Riazuelo IAP/UPMC/CNRS

**Figure 7.** Gravitational lensing by a black hole in the line of sight of the Magellanic Clouds. The large rings are two mirror images of the Large Magellanic Cloud. The brightest star is Canopus, seen twice. Credit A. Riazuelo IAP/UPMC/CNRS

### 3. Numerical Astrophysics
Numerical simulations play a vital role in all the fields of astrophysics, partly because many astronomical phenomena are results of nonlinearly coupled multiple factors. Thus they are powerful tools to elucidate the properties and rules of such nonlinear systems. They also serve as numerical experiments, since real experiments are impossible in astrophysics. Thanks to rapid increase in computational power, numerical simulations have become sophisticated enough to achieve high spatial resolution, wide dynamic range and inclusion of many factors and effects. Such a sophistication enables us to compare the simulations with observations in wide areas of astrophysics. We give below two examples representative of the field.

*3.1. Simulations of type II supernovae explosions*
Type II supernovae result from the rapid collapse and violent explosion of stars more massive than 8 solar masses. They are among the most powerful cosmic phenomena, releasing more energy within seconds than stars like the sun produce in billions of years. They terminate the lives of massive stars, giving birth to neutron stars or stellar-mass black holes, and are responsible for the production of about half of the chemical elements heavier than iron. By far most of the energy release occurs in the form of elusive neutrinos, elementary particles that are abundantly produced when the hot, compact remnant settles and cools.

The major unsolved problem with Type II supernovae is that it is not understood how the burst of neutrinos transfers its energy to the rest of the star, producing the shock wave which causes the star to explode. Only one percent of the energy needs to be transferred to produce an explosion, but explaining how that one percent of transfer occurs has proven very difficult, even though the particle interactions involved are believed to be well understood.

A number of problems have plagued numerical simulations of type II supernovae for years. Supernova equations don't explode in 1D. One of the reasons is that stellar explosions are generically multi-dimensional phenomena. All implications of this fact were recognized only as computational models of supernovae explosions became increasingly more sophisticated and began to be advanced to the third dimension and to longer evolution periods. The other crucial area of investigation is the hydrodynamics of the plasma that makes up the dying star; how it behaves during the core collapse determines when and how the shock wave forms and when and how it stalls and is reenergized. Computer models are now successful at calculating the behavior of Type II supernovae [8][9]. They have led to a better understanding of the role of neutrinos for driving the explosion, to new insights into the characteristics of the neutrino and gravitational-wave emission of supernovae, and to the discovery of links between pulsar kicks and spins, supernova nucleosynthesis, and the origin of asymmetries seen in the gaseous remnants of stellar explosions.

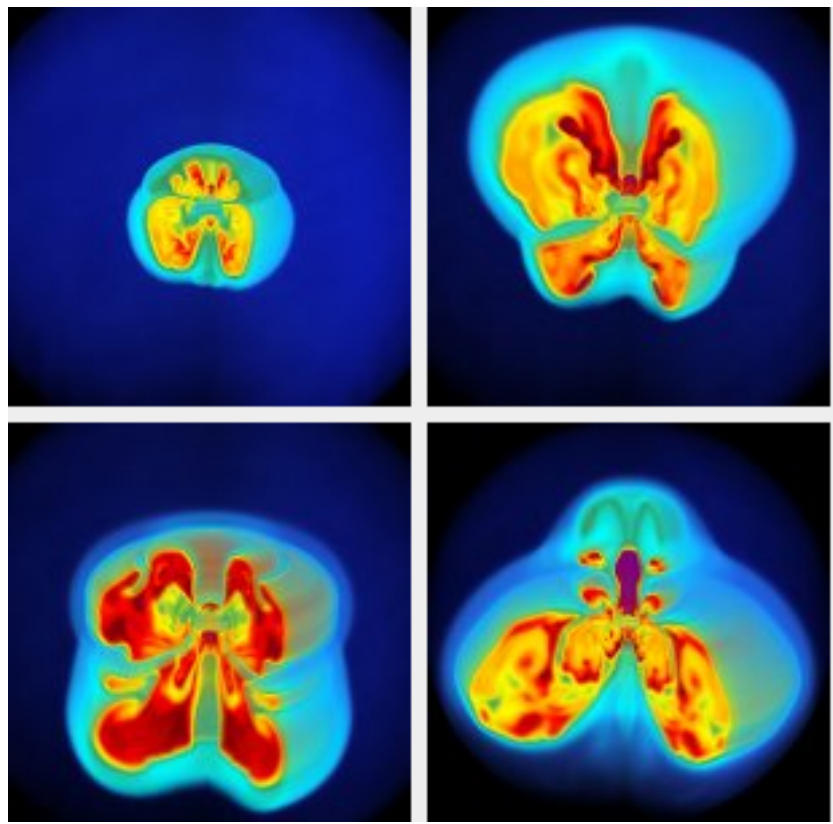

**Figure 8.** Onset of the type II supernova explosion of a star with 15 solar masses. The images show a sequence of moments from the computer simulation at 0.53, 0.61, 0.65, and 0.7 seconds (from top left to bottom right) after the stellar core has collapsed to a neutron star.

Full movie and more detailed explanations are available on http://www.youtube.com/watch?v=_ljcM4ftWnc (Visualization M. Rampp, RZG).

## 3.2. Tidal destruction of stars by giant black holes

Astrophysicists have long understood that massive black holes lurking in galactic nuclei and weighing millions of Suns can disrupt stars that come too close. Due to intense tidal forces, the black hole's gravity pulls harder on the nearest part of the star, an imbalance that pulls the star apart over a period of hours, once it gets inside the so-called "tidal radius". First calculations were performed by myself and collaborators in the 1980's [10]. We first used linearized equations to describe the final moments of a star as it penetrates deeply into the tidal field of a massive black hole, and we found that the tidal forces flatten it into a pancake shape; this flattening would increase the density and temperature inside the star enough to trigger intense nuclear reactions that would tear it apart.

But the real picture would be complicated by shock waves generated during the flattening process, so that other studies claimed that no nuclear explosion should occur. The controversy could be solved only by more sophisticated computer simulations, since no observational data were available. Various groups [11] [12] have recently investigated the effects of shock waves in detail; they confirmed that high compression and heating factors in the stellar core are able to trigger a thermonuclear explosion, and that the shock waves carry a brief but very high peak of temperature from the star's center outwards (figures 9-10), that could give rise to a new type of X-ray or gamma-ray bursts.

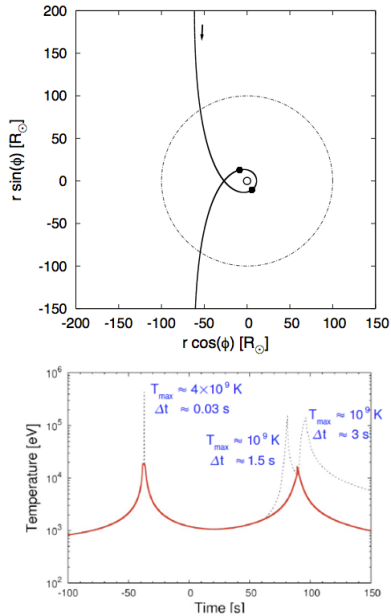 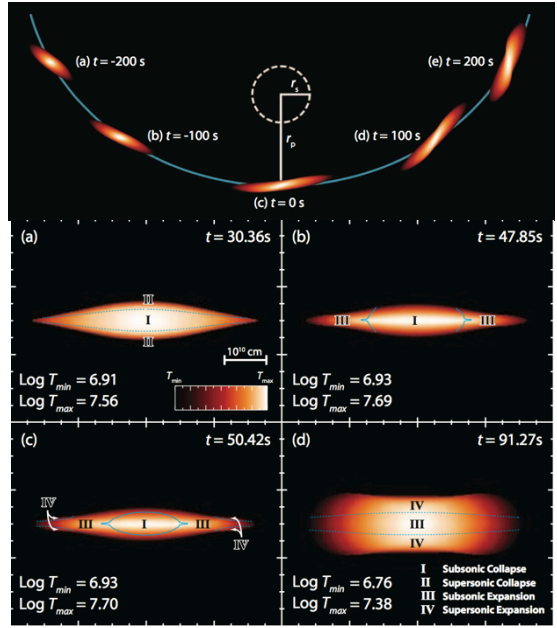

**Figure 9.** Top: Relativistic orbit of a solar-type star deeply plunging within the tidal radius of a $10^6$ M$_S$ black hole. The points on the orbit show the instants of maximum compression at the centre of the star. Bottom : as the stellar orbit winds up inside the tidal radius, the tidal field induces two successive compressions and heatings of the stellar core. 1D simulation. Credit: Luminet

**Figure 10.** Top : Tidal deformation of a star along a non-relativistic parabolic orbit around a $10^6$ M$_S$ black hole. Bottom : Four snapshots showing a slice through the center of mass perpendicular to the orbital plane and parallel to the minor axis of the remnant during the bounce phase of the passage. Regimes of subsonic/supersonic collapse/expansion are indicated with roman numerals. The 3D adaptive mesh grid code recovers the results of the 1D approximation. Credit : Guillochon et al.

Explosive disruption of stars by black holes will be an important subject in the next years. Recently observed X-ray, UV and gamma-ray bursts at the centers of galaxies are currently interpreted as the disruption of a star by the tidal forces exerted by the central massive black hole and place new constraints on the scenario that did not exist ten years ago. The search for X-ray flares, which would be due to stellar pancakes are now advancing to the next step, with new X-ray and gamma-ray observatories acquiring data. After the pioneering studies, the models have now to be improved with the inclusion of realistic equations of state for different stellar progenitors (solar-type, white dwarfs, red giant cores, ...) interacting with massive black holes in galaxies or intermediate-mass black holes in globular clusters, relativistic effects with Schwarzschild or Kerr metric (orbital precession induces multiple compressions), coupling with a nuclear reaction network, and computation of the energy transfer across the upper layers of the star in order to characterize the spectra of the resulting X-ray or gamma-ray flares. The aim is to predict observational signatures of these "tidally-induced supernovae" and to compare them to new data coming from telescopes.

**4. Numerical Cosmology**
In order to account for the observable Universe, any comprehensive model of cosmology must draw from many disciplines of physics, including general relativity, gauge theories of strong and weak interactions, the hydrodynamics and microphysics of baryonic matter, electromagnetic fields, spacetime curvature and topology, and so on. Although it is difficult to incorporate all these physical elements into a single complete model of our Universe, advances in computing methods and technologies have contributed significantly to our understanding of cosmological models.

*4.1. Dark Energy and Formation of Structures*
Here I closely follow [13]. 95 % of the cosmos is in the form of two invisible components, ~ 25 % is in Cold Dark Matter (CDM) particles, which are primarily responsible for the formation of the visible structures in the universe, ~ 70 % is in an unknown exotic form dubbed dark energy (DE), which is responsible for the present phase of cosmic accelerated expansion. The quest for DE has become one of the main topics of cosmological research. DE may have different theoretical explanations, but to date a coherent physical understanding is still missing. The study of DE through its imprints on cosmic structure formation is one of the core topics of investigation for numerical cosmology. One has to follow the gravitational collapse of dark matter throughout the history of the universe and across several orders of magnitude length scales for different models of DE. These calculations rely on N-body simulations. Many computational difficulties arise from the extreme number of particles involved, such as the great number of operations needed and the huge size of the generated data.

During the past 10 years several groups have pushed to the limits both size and resolution of cosmological N-body simulations. The Millenium Simulation in 2005 [14] has run a 2.2 billion light-years simulation box with 10 billion particles. Since then, the performance of cosmological simulations has increased more rapidly than the Moore's law (figure 11).

The Dark Energy Universe (DEUS) simulations [15] have performed the first-ever N-body simulations of the full observable universe (~95 billion light-years box) for three different DE models that are currently indistinguishable using cosmological observations. DEUSS have evolved 550 billion particles on an Adaptive Mesh Refinement grid with more than two and half trillion computing points along the entire evolutionary history of the universe and across 6 orders of magnitudes length scales, from the size of the Milky Way (mass of one particle) to that of the whole observable Universe. Such runs provide unique information on the formation and evolution of the largest structures in the universe and a support to future observational programs dedicated to the mapping of the distribution of matter and galaxies in the universe, such as the EUCLID mission (http://sci.esa.int/euclid). Each simulation has run on 4752 (of 5040) thin nodes of GENCI's supercomputer CURIE, using more than 300 TB of memory for 5 million hours of computing time. About 50 PBytes of rough data were generated throughout each run. Using an advanced reduction workflow, the amount of useful stored data has been reduced almost one the fly for each to 500 TBytes. Overall the realization of such large simulations required the development of a global application which integrated all aspects of the physical computational problem: initial conditions, dynamical computing, data validation, reduction and storage. Hence, it required optimizing not only the efficiency of the numerical dynamical solver, but also the memory usage, communications and I/O at the same time. Previous cosmological "grand-challenge" simulations could limit the use of efficient parallel HPC schemes to the dynamical computational solver only. Instead the DEUS experiment and similar projects clearly show the need for the use of MPI instructions in all parts of the application, scalable to more than 10,000 to 100 000 processes and optimized to match the specificities of supercomputing machine in which these have been run.

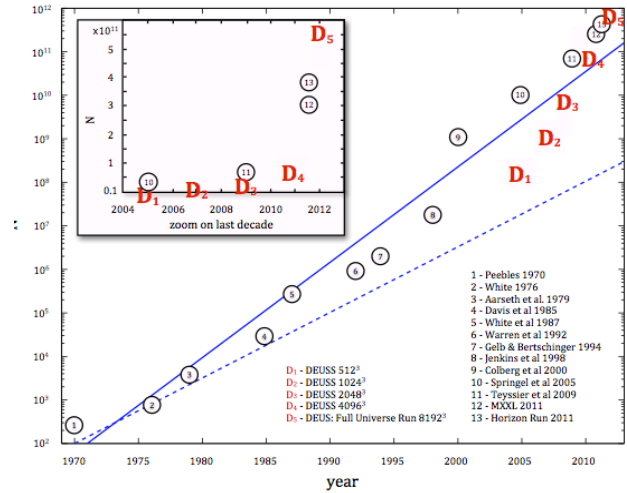

**Figure 11.** Evolution of cosmological N-body simulations over the last 40 years. A zoom shows the acceleration in performance occurred in the last decade, especially for DEUS collaboration. The solid line is the mean evolution of the simulation size, the dashed line is "Moore's Law", which corresponds to a factor 2 increase every 18 months. Courtesy J.-M. Alimi.

Computing time has been granted on a number of supercomputing facilities in France by GENCI. The realization of these simulations has had a great impact on the national supercomputing community in France : in fact to perform such project the supercomputing machines have been used at their full capacities. This has allowed testing and optimizing their nominal performances.

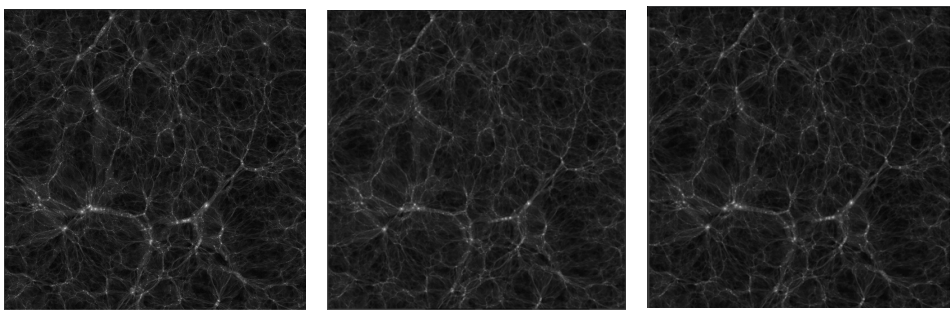

**Figure 12.** Formation of structures at the scale of 162 Mpc from DEUS simulations, in three different models of DE. Left : LCDM; center : RPCDM; right : SUCDM. Credit : DEUS-Consortium.org

Figure 12 depicts the formation of structures obtained by DEUS for three models of DE. Full movies are available on http://www.deus-consortium.org. The numerical data are publicly accessible through the Dark Energy Universe Virtual Observatory (DEUVO) database (http://roxxor.obspm.fr/deuvo-ui/).

*4.2. Cosmic Topology*

It is presently believed that our Universe is correctly described at large scale by a Friedmann-Lemaître (hereafter FL) model. The FL models are homogeneous and isotropic solutions of Einstein's equations, of which the spatial sections have constant curvature ; thus they fall into 3 general classes according to the sign of their spatial curvature, resp. k= +1, 0, or -1.

The spatial topology is usually assumed to be the same as that of the corresponding simply connected, universal covering space, resp. the hypersphere, Euclidean space or the 3D-hyperboloid, the first being finite and the other two infinite. However, there is no particular reason for space to have a simply connected topology: to a given FL metric element there are several possible topologies, and thus different models for the physical Universe [16]. For example, the hypertorus and familiar Euclidean space are locally identical, and relativistic cosmological models describe them with the same FL equations, even though the former is finite and the latter infinite ; in fact the multi-connected cosmological models share exactly the same kinematics and dynamics as the corresponding simply connected ones.

The global topology of the universe can be tested by studying the 3D distribution of discrete sources and the 2D fluctuations in the Cosmic Microwave Background (CMB). The methods are all based on the search for topological images of a same celestial object such as a galaxy, a cluster or a spot in the CMB. Such topological images can appear in a multi-connected space a characteristic length scale of which is smaller than the size of the observable space, because light emitted by a distant source can reach an observer along several null geodesics [17]. Figure 13 shows a computer visualization of the structure of the Poincaré Dodecahedral Space (PDS), a positively curved space which has been proposed [18] to account for the peculiarities of the observed CMB angular power spectrum.

If space has a non trivial topology, there must be particular correlations in the CMB, namely pairs of "matched circles" along which temperature fluctuations should be the same (figure 14). The PDS model predicts 6 pairs of antipodal circles with an angular radius less than 35°. Such circles have been searched in WMAP and PLANCK data by several teams, using various statistical indicators and massive computer calculations. The simplest multi-connected models, in which the matched circles are antipodal or nearly antipodal, such as the hypertorus and PDS, seem to be excluded by recent analyses [19].

However in generic topologies the circles are not antipodal, and their position in the sky depend on the observer's position in the fundamental domain. The corresponding larger number of degrees of freedom for the circle search in the data generates a dramatic increase of the computer time, up to values which are out-of-reach of the present facilities. To be more precise, a full search requires 6 free parameters : 2 for the location of first circle center, 2 for the location of second circle center, 1 for the radius of the circle and 1 for the relative phase of the two circles. A full search naively costs about $n^{5/2}$ operations, where n is the number of sky pixels. The recently released Planck CMB map has $5.10^7$ pixels, which implies $10^{27}$ operations, i.e. ~3,000 years of computing time on petaflop supercomputers! Various computational tricks can reduce the cost, but not much.

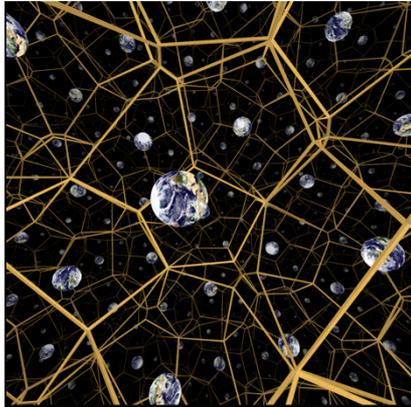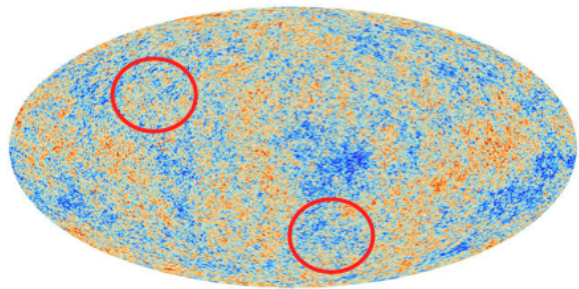

**Figure 13.** View from inside PDS along an arbitrary direction, calculated by the CurvedSpaces program [20], with multiple images of the Earth. The observer has the illusion to live in a space 120 times vaster, made of tiled dodecahedra which duplicate like in a mirror hall. Credit : J Weeks.

**Figure 14.** A pair of matched circles arbitrarily superimposed on the Planck CMB map. The number, location and size of matched circles depend on space topology and characteristic lengths of the fundamental domain compared to the size of the observable universe. A positive detection would be a signature of a multi-connected space at a sub-horizon scale. Credit : Planck Collaboration.